\documentclass[aps,prl,showpacs,nofootinbib,floats,floatfix,preprintnumbers,groupedaddress,twocolumn]{revtex4-1}
\usepackage{graphicx,epsfig}
\usepackage{dcolumn}
\usepackage{bm}
\usepackage{latexsym}
\usepackage{color}
\usepackage{tabularx}
\usepackage{ulem}
\usepackage{hyperref}
\usepackage{float}
\usepackage{tabularx}
\usepackage{color}
\usepackage{comment}
\usepackage{hyperref}
\usepackage{colortbl}
\usepackage{listings}
\usepackage{amsmath,amsfonts,amssymb}
\usepackage{fancyhdr}
\usepackage{hyperref}
\usepackage{natbib}
\hypersetup{
	colorlinks   = true, 
	urlcolor     = blue, 
	linkcolor    = blue, 
	citecolor    = red 
}

\hypersetup{
	colorlinks   = true, 
	urlcolor     = blue, 
	linkcolor    = blue, 
	citecolor   = red 
}

\begin{document}
\title{{Horizon induces instability locally and creates quantum thermality}} 
\author{Surojit Dalui}
\email{suroj176121013@iitg.ac.in}
\author{Bibhas Ranjan Majhi}
\email{bibhas.majhi@iitg.ac.in}
\author{Pankaj Mishra}
\email{pankaj.mishra@iitg.ac.in}
\affiliation{Department of Physics, Indian Institute of Technology Guwahati, Guwahati 781039, Assam, India}

\date{\today}

\begin{abstract}
	The classical Hamiltonian for a chargeless and massless particle in a very near horizon region is shown to be of the form $H\sim xp$ as long as radial motion is concerned. This is demonstrated explicitly for static spherically symmetric black hole and also found to be applicable for specific choice of radial trajectories in the Kerr case. Such feature of horizon leads to unavoidable ``{\it local instability}'' in the particle's radial motion as long as near horizon regime is concerned. We show that at the quantum level this provides thermality in the system. The temperature is found to be given by the Hawking expression. Finally, we conjecture that the automatic instability created by the horizon is responsible for its own temperature and consequently can be a possible physical mechanism for horizon temperature.
	
\end{abstract}

\maketitle

{\it Introduction. --}
Over the past few years, there has been an upsurge of interest in the study of the close relationship between the geometrical properties of horizons and the dynamics of particle motion near it. Recently, people have observed that horizons have some fascinating characteristics, one of which is its influence on integrable systems and sometimes turns it into a chaotic one, depending upon the the values of available  parameters (like energy of the particle, mass of the black hole, etc.). This has been established classically \cite{Cardoso:2008bp,Hashimoto:2016dfz,Hashimoto:2018fkb,Dalui:2018qqv,Lu:2018mpr,Zhao:2018wkl,Cubrovic:2019qee,Dalui:2019umw}. Its quantum consequence has also been studied to some extent. Usually, the quantum chaos can be diagnosed by the exponential increasing behavior of the out-of-order time correlation function (OTOC) of some quantum operators \cite{Maldacena:2015waa,Hashimoto:2017oit} and the size of them \cite{Susskind:2018tei,Brown:2018kvn}. In fact, one can relate the growth of the operator with the acceleration of the particle when it falls towards the black hole (BH). 

In all these analyses, it is found that the Lyapunov exponent (LE), which characterises the chaos in the system, has an upper bound as predicted by the  Sachdev-Ye-Kitaev (SYK) model \cite{Maldacena:2015waa}. For BHs it is predicted by the exponential growth of the radial motion of the particle \cite{Hashimoto:2016dfz,Dalui:2018qqv,Dalui:2019umw}. Interestingly, it has been noticed that such behavior persists for the static (e.g. Schwarzschild spacetime) as well as for the stationary (e.g. Kerr spacetime) BHs. In all cases the upper bound on the LE is determined by the surface gravity of them \cite{Dalui:2018qqv,Dalui:2019umw}. This universality is not so obvious and, as far as we know, the reason behind it has not been discussed anywhere. 

On the other hand, BH thermodynamics \cite{Bekenstein:1973ur, Bardeen:1973gs, Hawking:1974rv} is a long standing concept which originates through an {\it analogy} between the laws of BH and those of usual thermodynamical system, and it remains same till date. In fact, the underlying physical mechanism which provides temperature, is not illuminated till now. More precisely, how and why horizon gets temperature is not known to us. For instance, the source of temperature of a gas, contained in a cylinder, is due to the kinetic energy of the gas molecules. Similar mechanism in the case of horizon is not known.  Since maximum value of the LE is given by the surface gravity of black hole \cite{Maldacena:2015waa} and also the horizon temperature is determined by this one \cite{Hawking:1974rv}, the aforesaid instability due to horizon might give deep insight of understanding the mechanism of origination of temperature. Here we aim to enlighten all these issues which leads to a new concept. 

Sometimes, the local features of a system, although may be absent in a global sense, can give the possible explanation of the phenomenological behavior of the system. Taking up this idea here we investigate the behavior of a massless, chargeless particle in the vicinity of the black hole horizon which is moving {\it only radially outward direction very near to the horizon}. 
We find that near horizon Hamiltonian of the particle, at the leading order in radial distance from the horizon, is $H\sim xp$, where $p$ is the conjugate momentum of position variable $x$. We show this explicitly for any static spherically black hole. Moreover the same can be achieved for Kerr case as well for a particular trajectory. This form of near horizon Hamiltonian is very universal in the sense that it also appears for any static, spherically symmetric BH  as well as a particle in an accelerated frame. Therefore, such a universality may help us to answer the very basic and fundamental as well as long standing questions.  In fact, this Hamiltonian is shown to be that of an inverted harmonic oscillator (IHO) in a new set of canonical variables and, therefore, providing {\it instability} in the system. Note that this exists only very near to horizon and hence we call it as ``{\it local instability}''. 
We argue that such instability may be the cause for the chaotic dynamics in a system, which has been shown earlier in numerical calculations \cite{Dalui:2018qqv,Dalui:2019umw}. 

Quantum consequences of this instability are also investigated. We observe that the density of states in this case is thermal in nature and the temperature is given by the Hawking expression $T=\hbar\kappa/2\pi$, where $\kappa$ is the surface gravity \cite{Hawking:1974rv}. It is indeed the quantum response to the classical instability triggered by the horizon.  Therefore we feel that the unavoidable instability created by the horizon itself in its nearby region may not only create chaos in a system at the classical level, but also keeps its imprint by making the system thermal at the quantum level. Finally, we conjecture that {\it this ``local instability'' can be the source of the horizon temperature}, and hence may be regarded as a possible mechanism for feeling horizon as hot object. Interstingly, since results are related to IHO, these are equally valid for the systems with this Hamiltonian in different branches of physics.


{\it Hamiltonian. --}
Consider a static spherically symmetric BH. Since we are interested in physics near the horizon, in order to remove the coordinate singularity at the horizon, we shall express the metric in Painleve coordinates \cite{Painleve} {\footnote{In principle, near horizon physics can be studied in any well behaved coordinates around the horizon, like those found in \cite{Doran:1999gb}. But here we adopted Painleve coordinates as it is being extensively used from very early days to study horizon properties.}}. 
Note that we are interested in the near horizon physics. Here we consider only the radial motion of the particle. The same has already been used in several occasions to study the near horizon physics, like in investigation of Hawking effect using gravitational anomalies \cite{Robinson:2005pd,Iso:2006wa,Majhi:2011yi,Banerjee:2007qs,Banerjee:2008sn,Banerjee:2008wq} as well as tunneling formalism \cite{Parikh:1999mf,Banerjee:2008sn,Banerjee:2008cf,Banerjee:2009wb} (see also \cite{Chakraborty:2019ltu,Banerjee:2019tbr} for some recent works). 
In this case considering only the radial motion of the particle, we can obtain the form of the Hamiltonian as:
\begin{eqnarray}
H=p_{r}\left[1-\sqrt{1-2\kappa(r-r_{H})}\right]~,
\label{Energy_approx}
\end{eqnarray}
where $\kappa$ is the surface gravity of the black hole and $p_r$ is radial momentum corresponding to radial coordinate $r$. $r_H$ is the location of horizon. Details can be seen in \cite{Dalui:2018qqv}. Just to mention, the same can also be obtained for Kerr case by considering only $\theta=0$ and $\phi$ fixed trajectories, with
\begin{eqnarray}
\kappa=\left(r_{H}^{2}-a^{2}\right)/\left(2r_{H}(r_{H}^{2}+a^{2})\right)~,
\label{kappa}
\end{eqnarray}
where $a$ is the angular momentum per unit mass. For Kerr case, at this point the outer region of ergosphere and event horizon coincides and hence no ambiguity will arise. Details are given in \cite{Dalui:2018qqv}, particularly see Eq. (A11). So for stationary black hole the above argument is very restrictive.
Since we are interested near to the horizon, expanding (\ref{Energy_approx}) upto the first order one obtains
\begin{eqnarray}
H\simeq\kappa xp~,
\label{xp_hamiltonian}
\end{eqnarray}
where $x\equiv(r-r_{H})~~~\text{and}~~~p_{r}\equiv p$.
The same Hamiltonian can also be obtained for the particle motion in the near to Killing horizon regime in a Rindler frame \cite{Dalui:2019umw}. It shows a possible inherent property of the horizon, as long as near horizon dynamics of the massless particle is concerned \cite{State2}. 
Note that, this one is of the Berry-Keating type $H\sim xp$ \cite{Book1}, which provides instability into the motion of the particle dynamics.

{\it Classical regime. --}
We immediately see that the system, represented by (\ref{xp_hamiltonian}), has a hyperbolic point at $x=0$ and at $p=0$ which induces the instability into the particle's motion in the radial direction. This is of course there only in the vicinity of horizon. In particular, the solutions of the equations of motion
\begin{eqnarray}
\dot{x}=\kappa x,~~~~~\dot{p}=-\kappa p~,
\end{eqnarray}
with the dot refers to the derivative with respect to the {\it affine parameter}, say $\lambda$, which parametrises the path, 
are
\begin{eqnarray}
x(\lambda)=x(0)e^{\kappa \lambda}; \,\,\,\
p(\lambda)=p(0)e^{-\kappa \lambda}~.
\label{B10}
\end{eqnarray}
The above Eq. (\ref{B10}) shows that the radial motion is always unstable.

In order to see the time ($\lambda$ here) evolution of the neighbouring trajectories, consider a trajectory (say $l^{th}$) which starts at phase space position $(x_{A},p_A)$ at $\lambda_{A}$ and ends at $(x_{B},p_B)$ after one period $T_l=\lambda_B-\lambda_A$. We are interested to examine the effect when one changes the initial position and momentum slightly, say $(\delta x_A,\delta p_A)$. Use of (\ref{B10}) leads to the relation between the separations for $T_l>0$ as 
\begin{eqnarray}
\begin{bmatrix}
\delta x_{B}\\
\delta p_{B}
\end{bmatrix}
=\underbrace{\begin{bmatrix}
	e^{\kappa T_{l}} & 0\\
	0 & e^{-\kappa T_{l}}
	\end{bmatrix}}_{M_{BA}}
\begin{bmatrix}
\delta x_{A}\\
\delta p_{A}
\end{bmatrix}~.
\label{Monodromy_matrix}
\end{eqnarray}
The above matrix $M_{BA}$ is known as the Monodromy matrix \cite{Stockmann:1999}. We mention that only the classical periodic trajectories are important here and they contribute to the density of states at the quantum level \cite{Stockmann:1999}. However, for $H\sim xp$, the trajectories are unbounded in nature. In addition, here, $x$ is always positive (as the particle is moving outside the horizon), while $p$ can be both positive and negative. To satisfy periodicity condition, Berry-Keating used a particular type of boundary condition on $x$ and $p$ \cite{Book1}, which was later generalized by considering a phase space having fixed boundaries \cite{Sierra:2007vm}. But till now the most convincing existing prescription is to use a complexified version of $H$ in a new set of canonical variables which leads to a harmonic oscillator (HO) \cite{Private_Berry}. 
This later prescription will be used explicitly in the next stage where the semi-classical quantization of $H$ will be done.  Consequently, (\ref{Monodromy_matrix}) yields the homogeneous instability, represented by calculating the largest instability factor (IF) \cite{Strogatz}:
\begin{equation}
\lambda_{L,max}= \lim_{T_{l}\rightarrow \infty}\frac{1}{T_{l}}\ln\left[\frac{\delta x_{B}}{\delta x_{A}}\right]=\kappa>0~.
\end{equation}
Usually the positive instability factor is associated with chaos in the system.

We saw that the near horizon trajectories are the one which diverge exponentially. It explains why any integrable system must be affected when it reaches very near to the horizon. In particular, this local instability may influence chaotic dynamics to a system when it comes very near to the horizon for some particular values of available parameters. Such an event is consistent with the KAM theory \cite{nonlinear:02}. This can be a probable explanation for getting chaotic behaviour in the earlier numerical analysis \cite{Dalui:2018qqv} where a particle trapped in potential like harmonic one, kept under the BH background. Moreover, such a thing is happening irrespective of the consideration of spacetime. The universal property of the spacetimes is - all of them should contain a horizon. In addition, the presence of any intrinsic curvature of the spacetime is not crucial; only the existence of horizon alone is enough to make the motion of the particle chaotic. As for example, particle dynamics on Rindler metric in harmonic trap also exhibits chaotic behaviour \cite{Dalui:2019umw}. This feature of horizons is due to the fact that the leading order particle Hamiltonian, near horizon, is inherently $H\sim xp$, which provides the local unstable trajectories in the particle motion.

{\it Quantum consequences. --}
Apart from the classical perspective, the unique structure of the Hamiltonian $xp$ kind has a great consequence if we turn our attention to the quantum mechanics level. Particularly, we are interested in finding the response of the ``local instability'' at the quantum level. As this can manufacture instability, the usual quantization rule normally not applicable. In this regard, Gutzwiller's trace formula \cite{Gutzwiller:1990,Stockmann:1999} will be important one.

The density of states $\rho(E)$ for a particular energy $E$ is expressed in terms of Green function $G(q,q',t)$ as \cite{Gutzwiller:1990,Stockmann:1999} 
\begin{eqnarray}
\rho(E)&=&-\frac{1}{\pi}\text{Im}(\text{Tr}(G))
\nonumber
\\
&=&-\frac{1}{\pi}\text{Im}\left(\int G(q,q,E)dq\right)~,
\label{Density_of_states}
\end{eqnarray}        
where $q$ is the coordinate.
The trace of the Green function can be evaluated by the Gutzwiller's trace formula \cite{Stockmann:1999}:
\begin{eqnarray}
g(E)=\int dqG (q,q,E)&=&-\frac{i}{\hbar}\sum_{l}\frac{T_{l}}{\vert\vert M_{BA,l} -1\vert\vert^{\frac{1}{2}}}\times\nonumber\\
&&\exp\bigg[\frac{i}{\hbar}S_{l}(E)-i\frac{\mu_{l}\pi}{2}\bigg]~,
\label{Green_function}
\end{eqnarray}
where the summation is over all the classically allowed trajectories. In the above, $\mu_{l}$ is Maslov index for the $l$th trajectory and $S_{l}(E)$ is the Jacobi action
\begin{eqnarray}
S_{l}(q_{A},q_{B},E)=\int_{q_{A}}^{q_{B}} p dq~,
\label{B1}
\end{eqnarray}
calculated between two points $q_{A}$ and $q_{B}$.
$T_{l}$ is the period for the primitive orbit, which means the time needed for one passage and in terms of $S_{l}(E)$, it can be expressed as
\begin{eqnarray}
T_{l}=\frac{\partial S_{l}(E)}{\partial E}~.
\label{B5}
\end{eqnarray}
$M_{BA}$  in the denominator of (\ref{Green_function}) is our monodromy matrix (\ref{Monodromy_matrix}). $||\dots||$ stands for modulus of the determinant. The formula (\ref{Green_function}) is derived in path integral approach with the assumption that Hamiltonian can be expressed as $H=p^2/(2m)+V(x)$ where $m$ is the mass of the particle with momentum $p$, moving under the potential $V(x)$.

In our present analysis, $H$ is given by (\ref{xp_hamiltonian}). This can be casted in the required form by changing the variables from one canonical set to another \cite{Book1}:
\begin{equation}
x=\frac{1}{\sqrt{2}}(P-X); \,\,\,\
p=\frac{1}{\sqrt{2}}(P+X)~,
\end{equation}
so that one can use (\ref{Green_function}) in our case as well. In these new variables
(\ref{xp_hamiltonian}) becomes
\begin{eqnarray}
H=\frac{\kappa}{2}(P^2 - X^2)~.
\label{E_near_horiz_approx}
\end{eqnarray}
This implies that (\ref{xp_hamiltonian}) is simply a canonically rotated IHO. Comparing this with the usual form of Hamiltonian for a IHO: $H_{IHO}=\frac{P^{2}}{2m}-\frac{1}{2}m\omega^{2}X^{2}$, we found that for our system $m\equiv \frac{1}{\kappa}$ and $\omega\equiv \kappa$. 

In order to calculate $S_l(E)$ in (\ref{Green_function}) from (\ref{B1}) we will use the following procedure. A direct evaluation shows that the energy eigenvalues are that of HO with a naive substitution of the frequency $\omega\rightarrow i\omega_0$ \cite{Barton:1984ey,Gentilini,Bhaduri}. There after this substitution has been appeared to be very fruitful in the quantum description of the IHO. Since we are also interested to the quantum regime, the same prescription will be followed here.
Under $\omega\rightarrow i\omega_0$, we have $H_{IHO}\rightarrow H_{HO}=\frac{P^{2}}{2m}+\frac{1}{2}m\omega_0^{2}X^{2}$, which gives periodic motion in phase space. Now with this for a full periodic motion along the $l^{th}$ orbit, (\ref{B1}) yields the area in phase space under the curve with energy $E_l$. This is given by $S_l = (2\pi E_l)/\omega_0$. Therefore the analytic continued action for our system turns out to be
\begin{eqnarray}
S(E_{l})=-\frac{2\pi E_{l}}{i\kappa}~,
\label{B2}
\end{eqnarray}
where we have used $\omega_0\rightarrow -i\omega=-i\kappa$. Consequently, (\ref{B5}) yields $T_l=2i\pi/\kappa$.

It is well known that the path integration can be interpreted as the partition function when the time coordinate is complexified. The periodicity of complex time is identified as the inverse temperature. Using this idea Hawking showed that by Euclideanising the black hole metric the calculation of the partition function gives us correct expression for the entropy of the horizon. Here the periodicity of complexified time around horizon is identified as the inverse of the Hawking temperature \cite{Hawking:1979rv}.
In our calculation by complexifying the frequency we obtained $T_l=2i\pi/\kappa$. Interestingly, this value exactly matches with time period what Hawking found by Euclideanising the spacetime. Therefore, we feel that there may be a close connection between Hawking’s argument with our complexification of the frequency.

Therefore, substitution of this and (\ref{B2}) in (\ref{Green_function}) with $T_l\rightarrow iT_l$ we find
the expression for the density of states as
\begin{eqnarray}
\rho(E)=\frac{1}{\hbar \kappa}\sum_{l}\frac{1}{\sinh\pi}e^{-\frac{2\pi E_{l}}{\hbar\kappa}}\cos{\frac{\mu_{l}\pi}{2}}~.
\label{density_of_states_I.H.O}
\end{eqnarray}
Here in the denominator, we substituted
$\vert\vert M_{BA,l}-1\vert\vert ^\frac{1}{2} =  2\sinh(\kappa T_l/2)\rightarrow 2\sinh(i\kappa T_l/2)$ (see Eq. (\ref{Monodromy_matrix})).
The above one is thermal in nature and the temperature can be identified as
\begin{equation} 
T=\frac{\hbar\kappa}{2\pi}~.
\label{B3}
\end{equation}
Note that the particular form (\ref{density_of_states_I.H.O}) is the characteristic feature of  IHO which is an unstable system.  
Interestingly, this is identical to the Hawking's expression \cite{Hawking:1974rv} for horizon temperature. It may be pointed that IHO leads to thermal nature at the quantum level has also been reported recently in \cite{Morita:2019bfr, Hegde:2018xub, Maitra:2019eix}.

The Gutzwiller trace formula, for finding the total density of states (DOS) for an unstable system have two parts. One is the mean part, calculated for the action ($S_{l}(X_{A},X_{B},E)$) of vanishing path length; i.e. $|X_B-X_A|\rightarrow 0$ and another part corresponds to $|X_B-X_A|\neq 0$ (see section 7.3 and 8.1 of \cite{Stockmann:1999} for details), known as the oscillatory part.  Berry and Keating \cite{Book1} showed that in case of $xp$ kind Hamiltonian, the mean part of the counting function $<\mathcal{N}>$ can be calculated for a truncated case and from which the mean part of the DOS can be obtained using $<\rho(E)>=d<\mathcal{N}>/dE$. The asymptotic expression for this comes out to be positive. Our expression (\ref{density_of_states_I.H.O}) is oscillatory in nature as it has been obtained for $|X_B-X_A|\neq 0$. This can be both positive and negative, depending on the value of $\cos(\mu_l\pi/2)$. We would like to mention here that negative DOS is not at all surprising for near equilibrium systems. For instance, in literature \cite{Scarlatella:2019} the negative value of DOS has been reported for quantum system. In \cite{Gomez} it has been explained that DOS can have negative values when one has quasi-probability distribution, like Wigner distribution function, at the quantum level. This is mainly related to the situation when the system is little away from the equilibrium. DOS has a close relation with the Wigner function (see section 8.1.3 of \cite{Stockmann:1999}) and it has negative values for states which are classically not allowed. However, these negative values must vanish at the classical limit $\hbar\rightarrow 0$. One can check that this is exactly happening for (\ref{density_of_states_I.H.O}) as well. It is a pure quantum contribution and vanishes for $\hbar\rightarrow 0$, whereas the mean value is always positive. This fact is related to the unstable behaviour of the Hamiltonian. The significance of this, till date, is not well understood.


Now, earlier in the classical analysis we found that the IF has an upper bound: $\lambda_{L}\leq \kappa$. Therefore (\ref{B3}) yields 
$\lambda_{L}\leq \frac{2\pi T}{\hbar}$, which was conjectured earlier in SYK model \cite{Maldacena:2015waa}.
Consequently, one finds that the temperature of the system is bounded from lower as
\begin{eqnarray}
T\geq\frac{\hbar}{2\pi}\lambda_{L}~.
\label{B4}
\end{eqnarray}
In \cite{Hashimoto:2018fkb,Dalui:2018qqv,Dalui:2019umw}, the above inequality is obtained by using the classical prediction of the upper bound on LE with the assumption that the horizon has Hawking temperature. This temperature concept was taken as external information. But in the present analysis, we systematically derived this temperature, and so the above relation is now certain rather than prediction. This can also be predicted from the OTOC calculation which for large time $t$ yields $C(t)\sim e^{2\kappa t}$ \cite{State}. This is the signature of quantum property of chaos which also identifies $\lambda_{L,max}=\kappa$ and provides another way of defining largest IF.

This analysis implies that the ``local instability'', created by the horizon, may not only induce chaos in a system at the classical level; but also makes the system thermal at the quantum level by a minimum temperature. The unavoidable unstable environment in the near horizon region puts its automatic signature by making a system quantum mechanically thermal.  Here one must be careful that, the aforesaid near horizon instability does not mean the particle plus horizon system is chaotic one. Rather we are saying that this instability may lead to chaos in a system (e.g. particle trapped in a harmonic potential) under some certain circumstances. Interestingly, the quantum implication of this hyperbolic point is always emergence of temperature at the quantum level.

{\it Comments and outlook. --} 
In this paper, we successfully explored the possible causes of inducing chaotic dynamics by the horizon under certain circumstances. Here the static spherically symmetric black hole, Rindler case and very restrictive trajectories for Kerr black hole have been explored. It is observed that such is due to the appearance of $xp$ type Hamiltonian for a near horizon motion of a particle. Moreover, the consequences at the quantum level are found to be the automatic appearance of thermality in the system. The temperature has minimum non-zero value as long as there exists a horizon in the spacetime and $\lambda_L> 0$.

So it is evident that the presence of the horizon provides inherent ``local instability'' in the near-horizon region of spacetime. In other words, a very small spacetime region in the vicinity of the horizon, at the classical level, is always unstable as far as particle motion is concerned. This, at the quantum level, provides automatic thermality to the system. The temperature, here we found, is exactly given by the Hawking expression. Now if the collective system consists of the horizon and the particle, is closed and both of them are at thermal equilibrium, then this temperature can be associated as the horizon temperature. Therefore, we feel that the present discussion may unfold the deeper reason for having the temperature of the horizon at the quantum level. The local unstable trajectory, provided by the existence of the horizon, are the main source of quantum temperature in the system. Hence we claim that horizon itself always creates an unstable environment in its vicinity and when any object enters here feels temperature at the quantum regime. Thus we conjecture that {\it the automatic instability created by the horizon provides its own temperature}. Therefore we feel that the present analysis can  illuminate several near horizon physics both at the classical and quantum level. In fact, this instability can be a possible physical mechanism for creation of horizon thermality, which is, as far as we are aware of, not being addressed anywhere. Thus, the present analysis not only lead to a completely new avenue to understand black hole but also may provide a new paradigm in this area. Moreover, several features of IHO, which are being mentioned here, is also equally well applicable to systems where the basic Hamiltonian is of this kind. Therefore we feel that our present investigation has wide applicability in various branches of physics.

{\bf Acknowledgments:} We thank M. V. Berry for his useful comments on $xp$ Hamiltonian. We also thank the anonymous referee for several important comments which lead to considerable improvement of the manuscript.

\pagebreak

\end{document}